\documentclass{epl}

\usepackage{graphicx}
\usepackage{amsmath, bbm}
\usepackage{color}

\begin{document}

\title
{Discontinuity of the chemical potential in reduced-density-matrix-functional theory}
\shorttitle{Discontinuity of $\mu$ in RDMFT}

\author{N. Helbig\inst{1,2} \and N. N. Lathiotakis\inst{1} \and M. Albrecht\inst{3}
\and E. K. U. Gross\inst{1}}
\shortauthor{N. Helbig \etal}

\institute{
\inst{1} Institut f\"ur Theoretische Physik - Freie Universit\"at Berlin, 
Arnimallee 14, D-14195 Berlin, Germany\\
\inst{2} Fritz-Haber-Institut der Max-Planck-Gesellschaft, Faradayweg 4-6, 14195 Berlin,
Germany\\
\inst{3} Theoretical Chemistry FB 08 - University
of Siegen, 57068 Siegen, Germany}


\newcommand{\be}{\begin{equation}}
\newcommand{\ee}{\end{equation}}
\newcommand{\bea}{\begin{eqnarray}}
\newcommand{\eea}{\end{eqnarray}}
\def\v#1{\mbox{\boldmath $#1$}}

\maketitle

\begin{abstract}
We present a novel method for calculating the fundamental gap. To this end,
reduced-density-matrix-functional theory is generalized to fractional particle
number. For each fixed particle number, $M$, the total energy is minimized with
respect to the natural orbitals and their occupation numbers. This leads to a
function, $E_{\mathrm{tot}}^M$, whose derivative with respect to the particle
number has a discontinuity identical to the gap. In contrast to density
functional theory, the energy minimum is generally not a stationary point of the
total-energy functional. Numerical results, presented for alkali atoms,  the
LiH molecule, the periodic one-dimensional LiH chain, and solid Ne, are in
excellent agreement with CI calculations and/or experimental data.  
\end{abstract}

\section{Introduction}
Reduced-density-matrix-functional theory has recently attracted a lot of
attention \cite{MUE1984,BB2002,GU1998,CA2000,SS2002,HH2003,KH2004,LHG2005}. 
Functionals of the one-body reduced density matrix (1-RDM)
have been used very successfully in the calculation of correlation energies and
dissociation curves of small molecules \cite{GPB2005}. Given the success 
of these functionals 
it is interesting to evaluate their performance in the calculation of other
properties. One particularly interesting quantity in this context is the 
fundamental gap $\Delta$.
It is given by the difference between the ionization potential
and the electron affinity
\be\label{delta}
\Delta=I-A,
\ee
where
\bea 
\label{ip}
I &=& E_{\mathrm{tot}}^{N-1}-E_{\mathrm{tot}}^N,\\
\label{ea}
A &=& E_{\mathrm{tot}}^N-E_{\mathrm{tot}}^{N+1}. 
\eea  
Here, $E_{\mathrm{tot}}^N$ is the total ground-state energy of the neutral $N$
electron  system while $E_{\mathrm{tot}}^{N\pm1}$ are the ground-state energies
of a system with charge $\mp 1$. In the chemistry literature (see, e.g. Ref.
\cite{PY1989}), $\Delta/2$ is usually termed the {\it absolute hardness} of a
chemical species. Here, we use the term {\it fundamental gap} for both finite
and extended systems since the physical concept defined by Eq. (\ref{delta}) is
the same in both cases. A method to calculate ionization
potentials within reduced-density-matrix-functional theory (RDMFT) has 
recently been proposed \cite{PC2005}.

Within density functional theory (DFT) one can prove \cite{PPLB1982} 
that the fundamental gap is exactly
given by the orbital-energy difference $\Delta\epsilon$ between the lowest 
unoccupied and the highest occupied Kohn-Sham (KS) state plus a number,
$\Delta_{xc}$, which amounts  to the discontinuity of the exchange-correlation
potential upon adding and subtracting a fractional charge with respect to the
$N$-electron system. This discontinuity is zero for LDA and GGA \cite{PPLB1982}. 
Consequently, $\Delta\epsilon$ is the prediction for the gap within these 
approximations. This prediction, however, deviates strongly from the experimental 
values underestimating them by typically 30-50\%. Moreover, strongly 
correlated materials, like FeO and CoO are predicted by LSDA to be metals (zero gap) 
while, experimentally, these materials are anti-ferromagnetic insulators \cite{TOWK1984}. 
Within exact-exchange, one of the
variants of DFT functionals, the discontinuity $\Delta_{xc}$ differs from zero
leading to an overestimation of the fundamental gap. The results are very close
to those obtained within Hartree-Fock theory \cite{SMMVG1999}. Consequently, the
calculation of the fundamental gap  within DFT is still an open problem.

In the present article we propose a method for calculating the fundamental gap
by exploiting reduced-density-matrix-functional theory. We derive a
rigorous formula for the fundamental gap and give numerical results for finite
and extended systems.

\section{The Discontinuity of $\mu$ in RDMFT}
In RDMFT the one-body reduced density matrix
\be 
\label{densmat}
\gamma(\v r,\v r') = N\int d^3r_2...d^3r_N\Psi^*(\v r',\v r_2...\v r_N) 
\Psi(\v r,\v r_2...\v r_N)
\ee 
is used as the fundamental variable. Here, $\Psi(\v r_1...\v r_N)$ is
the many-body wave function of the interacting $N$-electron system. As was shown by
Gilbert \cite{G1975}, one can establish a rigorous one-to-one correspondence between the ground-state
wave function and the one-body density matrix. Therefore, all ground-state
observables are functionals of the 1-RDM. The main advantage of RDMFT
compared to DFT is that the kinetic energy as a functional of $\gamma(\v r,\v r')$ is
known exactly 
\be 
T[\gamma] = \int\int d^3r\: d^3r'
\delta(\v r-\v r')\left(-\frac{\nabla^2}{2}\right)\gamma(\v r,\v r'). 
\ee
Hence, writing the total energy in the form 
\be
E_{\mathrm{tot}}[\gamma]=T[\gamma]+E_{\mathrm{ext}}[\gamma]+E_{\mathrm H}[\gamma]
+E_{\mathrm{xc}}[\gamma], 
\ee 
where $E_{\mathrm{ext}}[\gamma]$ and $E_{\mathrm H}[\gamma]$ are the
usual external and Hartree energy functionals, respectively, leads to an
exchange-correlation energy which, in contrast to DFT, does not contain 
any kinetic-energy
contributions. 

A complication within RDMFT is the absence of a Kohn-Sham system: Due to the
idempotency of the density matrix of all non-interacting systems it is impossible
to reproduce the density matrix of an interacting system since the latter is
always non-idempotent. Nevertheless, one can directly minimize the total energy
with respect to the density matrix. In practice, this minimization is replaced by a
minimization with respect to the natural orbitals $\varphi_j$ and occupation numbers $n_j$ which are the
eigenfunctions and eigenvalues of $\gamma$
\be
\int d^3r'\gamma(\v r,\v r')\varphi_j(\v r')=n_j\varphi_j(\v r).
\ee
In the minimization, a set of boundary conditions 
for $n$'s and $\varphi$'s should be considered. For the natural orbitals the only condition
is orthonormality. For the occupation numbers there are two conditions which are necessary 
and sufficient for the ensemble $N$-representability of the one-body density matrix for integer 
particle number \cite{C1963}. They are 
the particle number conservation $\sum_{j=1}^{\infty} n_j = N$, and  
the condition $0 \leq n_j \leq 1$. The orbital orthonormality and the particle number 
conservation conditions 
can be implemented via the Lagrange multipliers $\mu$ and 
$\epsilon_{ij}$ and the functional to be minimized reads
\be\label{ffunc}
F[\gamma]=E_{\mathrm{tot}}[\gamma]-\mu\left(\sum_{j=1}^{\infty} n_j-N\right)
-\sum_{j,k=1}^{\infty}\epsilon_{jk}\left(\int d^3r
\varphi_j^*(\v r)\varphi_k(\v r)-\delta_{jk}\right).
\ee
The inequality condition $0 \leq n_j \leq 1$ allows for optimal
sets of $n$'s containing the border values zero and/or one. We refer to these 
states as pinned states. For these states the derivative $\delta F/\delta n_j$ is not equal to zero
in the range $[0,1]$ and one of the two borders, either zero or one, is accepted as the 
optimal value for $n_j$. This situation is most evident for
non-interacting particles, where $E_{\mathrm{tot}}$ is {\it linear} in all $n_j$'s.
The existence of an occupation number which is equal to one
means that the particular natural orbital exists in all the determinants 
with non-zero coefficient in the full CI expansion. This situation is rather exceptional for the exact 
wavefunction of a system of interacting particles. However, it is rather the rule than the 
exception for most of the existing approximate 1-RDM functionals for systems with more than two
electrons. These functionals have the tendency to produce occupation numbers equal to one for 
most of the core states. Consequently, 
the functional derivative of $F$ with respect to $\gamma$ does not vanish at the minimum energy.
One should mention that there are functionals which do not produce pinned states \cite{KOLLMAR2004}.
In DFT the situation is very different because any density which integrates to
the correct particle number is $N$-representable. Therefore, the functional
\be\label{Fdft}
F[\rho]=E_{\mathrm{tot}}[\rho]-\mu\left(\int d^3r \rho(\v r)-N\right)
\ee
always has a minimum with $\delta F/\delta \rho=0$ at the solution 
point so that $\delta E_{\mathrm{tot}}/\delta\rho(\v r)=\mu$.

In order to derive a formula for the fundamental gap within RDMFT we 
extend the definition of the total-energy functional $E_{\mathrm{tot}}[\gamma]$ to systems with
fractional particle number $M$. Such systems can be described as an
ensemble consisting of an $N$ and an $N+1$ particle state for $N\leq M\leq N+1$. The
resulting ensemble 1-RDM for the fractional particle number $M=N+\eta$
($N\in\mathbbm{N}$, $0\leq\eta\leq 1$) is given by 
\be
\label{eq:ensemble}
\gamma^M(\v r,\v r')=(1-\eta)\gamma^N(\v r,\v r')
+\eta\gamma^{N+1}(\v r,\v r'),
\ee
and the lowest energy of the ensemble is
\be
E_{\mathrm{tot}}^M=(1-\eta)E_{\mathrm{tot}}^N+\eta E_{\mathrm{tot}}^{N+1},
\ee
where $E_{\mathrm{tot}}^N$ and $E_{\mathrm{tot}}^{N+1}$ are the ground-state energies of the $N$
and $N+1$ particle system, respectively.
It can be shown  that, in complete analogy to the case of integer $N$,
the necessary and sufficient conditions for a given $\gamma$ to be 
decomposed as in Eq.~(\ref{eq:ensemble}) are: $\sum_{j=1}^{\infty} n_j = M$ and $0 \leq n_j \leq 1$, where
$n_j$ are the eigenvalues of $\gamma^M$ \cite{SHLDG2006}. 

\begin{figure}[t]
\begin{center}
\includegraphics[width=0.3\textwidth, clip]{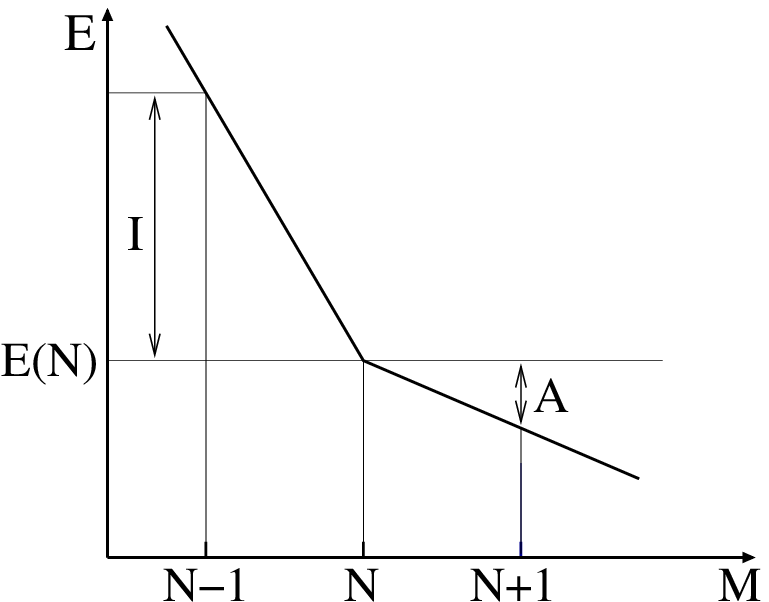}
\end{center}
\caption{\label{eversuseta} Total energy for fractional total number of particles.}
\end{figure}

As one can see in Fig.
\ref{eversuseta}, the derivative $\partial
E_{\mathrm{tot}}^M/\partial M$ has a discontinuity at the integer particle number $N$,
which by Eqs.~(\ref{delta}-\ref{ea}) is identical to the fundamental gap
\be\label{fundgap}
\Delta=\frac{\partial E_{\mathrm{tot}}^M}{\partial M}\biggl|_{N+\eta}
-\frac{\partial E_{\mathrm{tot}}^M}{\partial M}\biggl|_{N-\eta}.
\ee
So far, the derivation followed exactly the steps known from the 
generalization of DFT to fractional particle number \cite{PPLB1982}.
In the following we prove that the Lagrange multiplier, $\mu$, in (\ref{ffunc}) is
identical with the chemical potential, i.e.
\be\label{chempot}
\mu(M)=\frac{\partial E_{\mathrm{tot}}^M}{\partial M}.
\ee
In DFT, Eq. (\ref{chempot}) is a trivial consequence of the variational 
equation $\delta F/\delta\rho=0$ which implies that $\delta 
E_{\mathrm{tot}}/\delta\rho=\mu$ \cite{DG1990}. Consequently, 
$E_{\mathrm{tot}}^{M+\eta}-E_{\mathrm{tot}}^M=E[\rho^{M+\eta}]-E[\rho^M]=
\int d^3r\,\, \delta E/\delta\rho(\v r)|_{\rho^M}
(\rho^{M+\eta}(\v r)-\rho^M(\v r))=
\mu(M)\int d^3r \,\,(\rho^{M+\eta}(\v r)-\rho^M(\v r))=\mu(M)\:\eta$.
In our context of RDMFT however, this equivalence
is not at all obvious because $\delta F/\delta\gamma$ need not vanish at the
minimum energy (due to the states pinned at the border).
To prove (\ref{chempot}), we investigate the difference 
\be\label{difference}
E_{\mathrm{tot}}^{M+\eta}-E_{\mathrm{tot}}^M
 = E[\gamma^{M+\eta}]-E[\gamma^M]
=\int\int d^3rd^3r' \frac{\delta
E_{\mathrm{tot}}}{\delta\gamma(\v r,\v r')}\biggl|_{\gamma^M}
\left(\gamma^{M+\eta}(\v r,\v r')-\gamma^M(\v r,\v r')\right).
\ee
Using the fact that
\bea\label{deltan}
\frac{\delta n_j}{\delta\gamma(\v r,\v r')}&=&
\varphi_j^*(\v r)\varphi_j(\v r'),\\
\label{deltaphi}
\frac{\delta \varphi_j(\v x)}{\delta\gamma(\v r,\v r')}&=&
\sum_{\stackrel{\scriptstyle{k=1}}{k\neq j}}^{\infty}
\frac{\varphi_k^*(\v r)\varphi_j(\v r')}{n_j - n_k}\varphi_k(\v x)
\eea
the functional derivative of (\ref{ffunc}) yields
\be\label{fderiv}
\frac{\delta F}{\delta\gamma(\v r,\v r')}=
\frac{\delta E_{\mathrm{tot}}}{\delta\gamma(\v r,\v r')}
-\mu\delta(\v r-\v r').
\ee
Evaluating $\delta F/\delta\gamma$ via the functional chain rule, (\ref{fderiv})
leads to
\be
\frac{\delta E_{\mathrm{tot}}}{\delta\gamma(\v r,\v r')}=
\mu\delta(\v r-\v r')
+\sum_j \frac{\delta F}{\delta n_j}
\frac{\delta n_j}{\delta\gamma(\v r,\v r')}
+\sum_j\int d^3x \frac{\delta F}{\delta \varphi_j(\v x)}
\frac{\delta \varphi_j(\v x)}{\delta\gamma(\v r,\v r')}
+c.c.
\ee
{\it At the energy minimum}, $\delta F/\delta \varphi_j=0$ and $\delta F/\delta
n_l=0$ for any unpinned state $l$. The pinned states, however, contribute so
that
\be
\frac{\delta
E_{\mathrm{tot}}}{\delta\gamma(\v r,\v r')}\biggl|_{\mathrm{min}}=
\mu\delta(\v r-\v r')
+\sum_p \frac{\delta F}{\delta n_p} \varphi_p^*(\v r)\varphi_p(\v r'),
\ee
where we have used (\ref{deltan}) and the sum runs over those pinned states for
which $\delta F/\delta n_p\neq 0$.
Equation (\ref{difference}) therefore reduces to
\be
E_{\mathrm{tot}}^{M+\eta}-E_{\mathrm{tot}}^M=
\mu(M)\eta +\sum_p \frac{\delta F}{\delta n_p^M}
\times \int\int d^3rd^3r'
\varphi_p^{*M}(\v r)\varphi_p^M(\v r')
\left(\gamma^{M+\eta}(\v r,\v r')-\gamma^M(\v r,\v r')\right).
\ee
If we write the occupation numbers and orbitals of the $M+\eta$ system as
\be
\label{eq:deltas}
n_j^{M+\eta} = n_j^M +\delta n_j,\quad
\varphi_j^{M+\eta} = \varphi_j^M + \delta\varphi_j
\ee
we can obtain the first order corrections of $\gamma^{M+\eta}$. The changes
$\delta n_j$ and $\delta\varphi_j$ in Eq.~(\ref{eq:deltas}) are not arbitrary. They are the
changes in the optimal $n_j$ and $\varphi_j$ if an infinitesimal charge $\eta$ is
added to the system. Using the
orthonormality of the natural orbitals we get
\be
E_{\mathrm{tot}}^{M+\eta}-E_{\mathrm{tot}}^M = \mu(M)\eta + \sum_p \frac{\delta
F}{\delta n_p^M}
\times\left[\delta n_p +n_p^M\int d^3r
\left(\varphi_p^M(\v r)\delta\varphi_p^*(\v r)+
\delta\varphi_p(\v r)\varphi_p^{*M}(\v r)\right)\right].
\ee
The second term in the square brackets is zero since the norm of the natural orbitals does
not change and the first term is zero because the sum runs only over
pinned states and for these $\delta n_p = 0$, i.e. no states get unpinned. 
The states that would most likely get unpinned are those where
the ``true'' energy minimum ($\delta F/\delta n=0$) lies at a small
distance outside the allowed interval. However, this distance is still finite and
the infinitesimal additional charge $\eta$ cannot move the true minimum inside the allowed interval. 
This completes the proof of~(\ref{chempot}). Hence, by~(\ref{fundgap}), we can evaluate the
fundamental gap from the discontinuity of the Lagrange multiplier $\mu(M)$
\be\label{gap}
\Delta=\lim_{\eta\rightarrow 0}[\mu(M+\eta)-\mu(M-\eta)].
\ee

\section{Results}
Equation (\ref{gap}) is of course proven for the exact functional. It is interesting to
test it using approximate functionals $E[\gamma]$. We employed 
the functional of Goedecker and Umrigar~\cite{GU1998} (GU)
which has the same structure as the M\"uller functional~\cite{MUE1984,BB2002} with the important 
difference that the self-interaction terms are explicitly removed.
Our implementation for atoms and molecules is based on a Gaussian basis set
expansion of the natural orbitals and uses the GAMESS~\cite{SBBEGJKMNSWDM1993} computer code to 
calculate the one- and two-electron integrals.
We minimize the total energy with respect to both the
natural orbitals and the occupation numbers employing a conjugate gradient scheme.
Since the $N$-representability conditions for fractional particle number are identical to
those for integer~\cite{SHLDG2006}, the generalization to fractional particle numbers is straightforward.
We obtained the fundamental gap for several small molecules using 20 to 30 natural
orbitals which were expanded in a cc-PVQZ atomic Gaussian basis set~\cite{D1989}.

Fig.~\ref{muhe} shows the result of the numerical calculations for the LiH
molecule using the GU functional. There is a step near $M=4$ which is sharp at the lower edge but
relatively smooth at the upper edge. The discontinuity of $\mu(M)$ is located at
a value slightly higher than $M=4$ precisely at the point where the highest
fractional occupation number crosses one and gets pinned. These features are due
to the approximate nature of the exchange-correlation energy. The exact
functional would show the discontinuity exactly at $M=4$. In order to extract a
numerical value for the gap from the graph in Fig.~\ref{muhe} we used the
intersection of the extrapolated curve $\mu(M)$ for $M>4$ and a vertical line at the
position of the jump. 

\begin{figure}[t]
\twofigures[width=0.45\textwidth]{mu_N_LiH.eps}{Ne_discontinuity.eps}
\caption{The chemical potential $\mu$ (in Hartree) as a function of particle
number $M$ for the LiH molecule.}
\label{muhe}
\caption{The chemical potential $\mu$ (in Hartree) as a function of particle number $M$ 
for solid Ne. The value of the gap is compared with HF, various DFT
calculations \cite{SDD2005} and with experiment \cite{PHSSSKJ1976}.}
\label{solidne}
\end{figure}

\begin{table}[t]
\setlength{\tabcolsep}{0.15cm}
\caption{\label{tablegap} The fundamental gap for some atoms and small molecules as well as
the LiH chain, and solid Ne from RDMFT compared to other calculations and experimental values, 
all gaps are given in Hartree. 
$^a$QCI from Ref. \cite{MOP1994},
$^b$from Ref. \cite{RS1985}, 
$^c$ionization potential from \cite{MOP1994}, Electron affinity from
\cite{GM2004}, 
$^d$CISD using the same basis set as in RDMFT, 
$^e$ionization potential from Ref.\cite{IW1975}, Electron affinity from 
\cite{SG2000}, 
$^f$LDA, with CRYSTAL code \cite{DCORS2005} and the same basis set, 
$^g$GGA, with CRYSTAL code \cite{DCORS2005} and the same basis set, 
$^h$from Ref. \cite{PHSSSKJ1976}.}
\begin{center}
\begin{tabular}{c|ccc}
& RDMFT & Other theoretical & Experiment\\ \hline\hline
Li & 0.18 & 0.175$^a$ & 0.175$^b$\\
Na & 0.18 & 0.169$^c$ & 0.169$^b$\\
LiH molecule & 0.27 & 0.286$^d$ & 0.271$^e$\\ 
LiH chain & 0.64 & 0.500$^f$, 0.509$^g$ & \\ 
Ne solid & 0.76 & 0.439$^f$, 0.546$^g$ &
0.797$^h$\\\hline\hline
\end{tabular}
\end{center}
\end{table}

The results for the fundamental gap of Li, Na, and LiH calculated with the GU functional
are given in Table~\ref{tablegap} and are in very good agreement with state of the art CI
calculations. They also agree very well with experimental data. Note that in the
context of DFT, these values are exceedingly difficult to calculate because,
within standard LDA/GGA-type functionals, the negative ions of such small
systems are not even bound.

For periodic systems the symmetry properties of the many-body wave function 
imply that 
\begin{equation}
\gamma (\v r + \v R , \v r' + \v R ) = \gamma (\v r , \v r' ) 
\end{equation}
for arbitrary lattice vectors $\v R$. This property, on the other hand,
implies that the eigenfunctions of $\gamma$, i.e. the natural orbitals,
are Bloch functions, $\varphi_{\lambda \v k}(\v r)$, where $\lambda$ is a band index and 
$\v k$ is a wavevector in the first Brillouin zone~\cite{KG2001}. Hence, the spectral
representation of $\gamma$ reads
\begin{equation}
\label{eq:beta}
\gamma (\v r , \v r' ) = \sum_{\lambda, \v k} 
n_{\lambda \v k} \: \varphi^*_{\lambda \v k}(\v r') \:  \varphi_{\lambda \v k}(\v r)\,.
\end{equation}

In principle, one should now minimize the total energy with respect to the occupation numbers 
$n_{\lambda \v k}  $ and the natural orbitals $\varphi_{\lambda \v k}(\v r)$ as
described above for finite systems. However, with the approximate functionals
currently available, we encounter a serious difficulty: For the M\"uller functional 
$\mu (M) $ does not show any discontinuity for all the systems we studied 
so far. The self-interaction correction of
Goedecker and Umrigar\cite{GU1998} seems to be essential to reproduce this feature. 
However, in terms of Bloch orbitals, the self-interaction terms go to zero, i.e.,
the GU functional reduces to the M\"uller functional if Bloch orbitals are
inserted. To properly subtract the spurious self-interaction one has to use localized
orbitals\cite{STW1990,SG1988}, such as Wannier functions. Inserting 
the standard transformation from Bloch to Wannier functions, 
\begin{equation}
\varphi_{\lambda, \v k}(\v r) = \sum_{\v R} e^{i\v k \v R}\: \omega_{\lambda}(\v r - \v R) \,,
\end{equation}
in Eq.~(\ref{eq:beta}), the 1-RDM can be represented as
\begin{equation}
\gamma(\v r, \v r') = \sum_\lambda \sum_{\v R, \v R'} g_\lambda (\v R - \v R') \;
 \omega^*_{\lambda}(\v r' - \v R')\; \omega_{\lambda}(\v r - \v R)\,,
\label{eq:gam_wan}
\end{equation}
where $\omega_{\lambda}(\v r)$ is the Wannier function referring to band $\lambda$ and
\begin{equation}
\label{eq:E}
g_\lambda (\v R - \v R' ) = \sum_{\v k} n_{\lambda \v k} \: e^{i \v k (\v R - \v R')} \,.
\end{equation}
For systems with a gap, the Wannier functions decay exponentially. Hence, we expect that
the products $\omega^*_{\lambda}(\v r' - \v R')\: \omega_{\lambda}(\v r - \v R) $ contribute
very little to $\gamma$ if $\v R \neq \v R'$. As a first implementation we neglect 
these off-diagonal terms altogether by making the approximation 
\begin{equation}
g_\lambda (\v R - \v R' )= n_\lambda \; \delta(\v R - \v R' )
\label{eq:F}
\end{equation}
which leads to
\begin{equation}
 \gamma(\v r, \v r') = \sum_{\lambda, \v R} n_\lambda \; 
 \omega^*_{\lambda}(\v r' - \v R) \; \omega_{\lambda}(\v r - \v R) \,.
\label{eq:G}
\end{equation}
Restricting the search to density matrices of the form~(\ref{eq:G}), we can then
go ahead and minimize the total energy with respect to the Wannier functions 
and their occupation numbers $n_\lambda$ using the GU functional. 
The self interaction terms, when evaluated with Wannier orbitals, do not vanish and we obtain reasonable 
results (see below). The restricted search over density matrices of the form (\ref{eq:G}) can be
viewed in yet another way: By Eq.~(\ref{eq:E}), the approximation (\ref{eq:F}) amounts to neglecting
the k-dependence of the Bloch occupation numbers, $n_{\lambda \v k} \approx n_{\lambda} $. This 
is expected to be a good approximation for insulators. For metals on the other hand, the 
approximation breaks down completely because $n_{\lambda \v k}$ changes, at the Fermi surface,
from values close to one to values close to zero within the same band. 

We implemented the minimization of the 1-RDM functionals in the space of Wannier states using
the Wannier computer code described in~\cite{SDFS1998,SDS1998}. As a first test case,  
we considered a system in one dimension, namely the LiH chain.  
Like in the case of finite systems, $\mu(M)$ shows a pronounced step. The size of this
step compares favorably with the LDA and GGA values (see Table~\ref{tablegap}).
Clearly, there are no experimental data available for this system but, as
always, the LDA/GGA results are expected to be smaller than the true value. 

As a first fully three-dimensional system, we performed a calculation for solid
Ne. Fig.~\ref{solidne} shows the discontinuity of the chemical potential when
only the occupation numbers are optimized. The discontinuity in Ne appears
slightly above 10 which is again due to the approximate nature of the
exchange-correlation functional. The value of the gap, extracted from Fig.~\ref{solidne} 
by extrapolation, is also included in Table~\ref{tablegap} and
compares very well with the experimental value. 

In conclusion, we have presented a method to calculate the fundamental gap of
finite systems and periodic solids within reduced-density-matrix-functional
theory. First numerical results were obtained using a recently proposed 1-RDM
functional. For all systems studied, the chemical
potential shows a clear discontinuity as a function of the total number of
electrons if all self-interaction terms are removed. The extracted gap 
values agree better with CI calculations and/or
the experiment than any standard DFT or Hartree-Fock calculation. 

\acknowledgments
This work was supported by the Deutsche Forschungsgemeinschaft (program SPP 1145),
by the EXCITING Research and Training Network and by the NANOQUANTA Network of Excellence.

\end{document}